# Extended Spherical Geometry Algorithm for Spaceborne SAR Processing in Stripmap and TOPS Imaging Modes

Xinhua Mao,  Manyi Tao,  Jixia Fan

*Abstract*—The Spherical Geometry Algorithm (SGA) demonstrates superior capability in achieving efficient and precise spaceborne SAR image formation processing, even under challenging imaging conditions including non-linear radar trajectories and spherical Earth surface geometry. Nevertheless, the original SGA is specifically developed for spotlight SAR data processing and can't directly applied to processing spaceborne SAR data in other modes. In this paper, we first analyze the limitations of the SGA algorithm when applied to stripmap or TOPS mode SAR processing, and then propose an improved SGA algorithm which can process both stripmap and TOPS SAR data. Compared with the original algorithm, the new algorithm has two main differences. Firstly, in order to avoid undersampling during azimuth resampling in both modes, an instantaneous Doppler centroid removal process was added before azimuth interpolation processing by exploiting the endomorphism property of resampling operation. Secondly, the spectral analysis method used for the final step of azimuth compression in the original SGA has been replaced with a new matched filtering processing, which can avoid image aliasing in azimuth direction and improve computational efficiency. Measured real data processing results are presented to demonstrate the validity of the proposed algorithms.

## I. INTRODUCTION

SPACEBORN Synthetic Aperture Radar (SAR) can provide high-resolution observations of the Earth's surface at all times and in any weather condition, offering many advantages that traditional optical and infrared sensors cannot provide [1-3]. However, SAR remote sensing also has its disadvantages compared to optical and infrared methods. The most significant drawback is that SAR imaging requires complicated signal processing to produce high-resolution images after data acquisition [4]. After decades of development, SAR imaging technology has gradually matured.

However, traditional high-efficient space-borne SAR image formation algorithms all rely on two assumptions: a uniform linear radar flight trajectory and a flat imaging scene [5]. These assumptions generally hold true for traditional low-resolution spaceborne SAR systems on low-orbit satellites. However, with the continuous improvement of the resolution of space-borne SAR imaging in recent years, the synthetic aperture time gradually reaches several seconds or even tens of seconds [6-8]. Over such a long synthetic aperture time, it is clear that the assumption of linear motion of the radar platform is often no longer valid and that satellite orbit curvature effect has to be considered [9-11]. To make matters worse, the Earth rotation effect causes the satellite trajectory relative to the ground scene become a more complex, non-coplanar curve. On the other hand, as the dimension of imaging scenarios increase [12-14], the planar ground surface assumption that traditional imaging algorithms relied on will no longer be applicable, and precise image formation processing will have to take into account the spherical ground surface effect. For the nonlinear radar flight trajectory problem, most of the current solutions are to approximate the actual range history between the radar and the scatterers by increasing the order of the range history model [15-19]. However, it is still difficult to meet the accuracy requirement when the resolution becomes very high, and on the other hand, it is also difficult to establish a uniform and accurate range history model for different scatterers in the illuminated scene. For the spherical ground surface problem, the traditional strategy is to regard it as a detrimental factor and compensate for the additional space-variant phase error introduced by the flat ground surface assumption through the addition of a complicated space-variant phase correction process. Therefore, it is difficult for these methods to balance both algorithmic accuracy and computational efficiency simultaneously.

The pioneering work in Reference [20] introduces an groundbreaking approach that strategically converts the inherent limitations of spherical Earth geometry into a computational advantage. By strategically utilizing the spherical surface effect, this method establishes an exact Fourier transform relationship between echo data and scene reflectivity, thereby achieving precise SAR imaging of the spherical surface under nonlinear radar trajectories. The new algorithm has no limitation on the imaging scene size and resolution, so it can be theoretically applied to any SAR imaging modes. However, the original SGA algorithm is

This work was supported in part by the National Natural Science Foundation of China under Grant 62471227 and SAST Foundation under Grant SAST2023-029. *(Corresponding author: Xinhua Mao, Jixia Fan).*

Xinhua Mao is with the College of Electronic and Information Engineering, Nanjing University of Aeronautics and Astronautics, Nanjing 211106, China (e-mail: xinhua@nuaa.edu.cn).
Manyi Tao is with Shanghai Institute of Satellite Engineering, China.
Jixia Fan is with Shanghai Academy of Spaceflight Technology, Shanghai, 201109, China (13701609177@163.com)

proposed only for the spotlight mode data, which cannot be directly applied to other imaging modes, such as stripmap mode or TOPS mode, due to the sampling ambiguity. In this study, we first conduct a detailed analysis of Doppler frequency history variations during SGA processing under ideal sampling conditions. Subsequently, we systematically examine the sampling ambiguity challenges encountered when applying the original SGA algorithm to stripmap and TOPS SAR data processing. Building upon these investigations, we propose an enhanced SGA framework specifically optimized for spaceborne SAR data acquisition in stripmap and TOPS imaging modes. Compared with the original SGA, there are two main differences in the improved algorithm. First, a dedicated phase adjustment module is incorporated prior to azimuth resampling, ensuring alignment between the total azimuth signal bandwidth and the instantaneous bandwidth requirement. This critical modification can guarantee mathematically rigorous azimuth interpolation. Secondly, the spectral analysis method employed in the final azimuth compression stage has been replaced by an advanced matched filtering scheme. This innovation effectively eliminates azimuthal image aliasing artifacts in azimuth direction and improve computational efficiency.

The rest of the paper is organized as follows. In section II, the standard SGA for spotlight SAR is briefly introduced for the sake of completeness. Then section III discusses the undersampling problem of the SGA when processing the stripmap and TOPS SAR data and presents an extended SGA for processing data collected in both modes. Finally, in section IV, experimental results are presented to demonstrate the effectiveness of the proposed approach.

## II. SPHERICAL GEOMETRY ALGORITHM FOR SPOTLIGHT SAR

This section briefly introduces the theoretical framework of the traditional SGA for spotlight SAR data processing, laying the foundation for subsequent analysis of time-frequency characteristics in range history across various SGA processing stages under different SAR operational modes. The original SGA implementation exists in two distinct variants: a basic formulation neglecting non-coplanar effects induced by Earth's rotation, and an enhanced version incorporating these geometric considerations. For analytical simplicity and conceptual clarity, our current investigation focuses exclusively on the former scenario, while recognizing that the proposed methodology can be directly extended to the latter case.

### A. Echo Model

Assuming the data collection geometry of spaceborne SAR is shown in the Fig.1. A Cartesian coordinate system is established with the center of the Earth as the coordinate origin, the radar position vector at the aperture center as the $Y$ axis, the radar velocity direction as the $X$ axis, and the $Z$ axis is defined following the right-hand rule. In this coordinate system, the instantaneous position of the radar is denoted as $(R\sin\theta, R\cos\theta, 0)$, where $R$ and $\theta$ are the instantaneous slant range and azimuth angle of the radar relative to the center of the Earth, respectively. Assuming an ideal point target is located on the surface of the Earth, its position is $(x_t, y_t, z_t)$.

Assuming that the radar transmits a wideband RF signal, the received echo signal from the point target after demodulation and pulse compression processing can be expressed as

$$S(t_a, t_r) = \text{rect}\left(\frac{t_a}{T_a}\right) \cdot \text{sinc}\left(B_r\left(t_r - \frac{2r}{c}\right)\right) \cdot \exp\left\{-j\frac{4\pi f_c}{c}r\right\}, \quad (1)$$

where $t_a$ and $t_r$ are the azimuth time and range time, respectively, $T_a$ is the synthetic aperture time, $c$ is the speed of electromagnetic wave, $f_c$ and $B_r$ are the carrier frequency and bandwidth of transmitted signal, respectively, $r$ is the range history determined by

$$r(t) = \sqrt{(R\sin\theta - x_t)^2 + (R\cos\theta - y_t)^2 + (z_t)^2} . \quad (2)$$

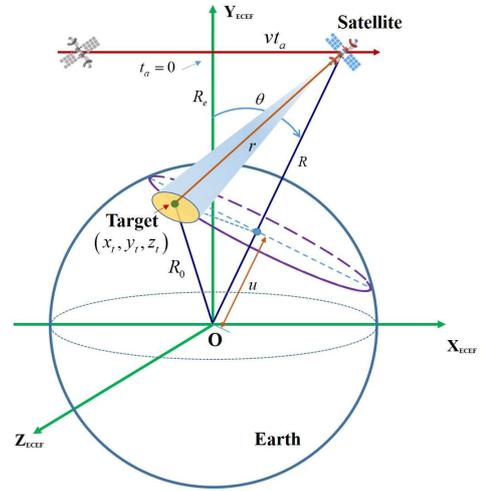

Fig.1. Data collection geometry of spaceborne SAR.

### B. Range Preprocessing

The first step in the SGA is range preprocessing, which includes range resampling, phase compensation, and range Fourier transform. The range resampling operation can be mathematically described by a change-of-variable in range dimension as follows

$$t_r = \zeta(\bar{t}_r) = \frac{2}{c}\sqrt{R^2 + R_0^2 - Rc\bar{t}_r} , \quad (3)$$

where $R_0$ is the radius of the Earth and $\bar{t}_r$ is the new range time variable after change-of-variable processing.

The signal after range resampling can be expressed as

$$S(t_a, \bar{t}_r) = \text{rect}\left(\frac{t_a}{T_a}\right) \cdot \text{sinc}\left(\bar{B}_r\left(\bar{t}_r - \frac{2u}{c}\right)\right) \cdot \exp\left\{-j\frac{4\pi f_c}{c}r\right\} , \quad (4)$$

where $u = x_t \sin\theta + y_t \cos\theta$, $\bar{B}_r = -\frac{R}{r}B_r$.

The second step is phase compensation. The phase compensation function is

$$H(\bar{t}_r) = \cdot \exp\left\{j\frac{4\pi}{c}(f_c r - \bar{f}_c u)\right\}, \quad (5)$$

where $\bar{f}_c = -\frac{R}{r} f_c$.

After phase correction, the signal becomes

$$S(t_a, \bar{t}_r) = \text{rect}\left(\frac{t_a}{T_a}\right) \cdot \text{sinc}\left(\bar{B}_r\left(\bar{t}_r - \frac{2u}{c}\right)\right) \cdot \exp\left\{-j\frac{4\pi \bar{f}_c}{c} u\right\}. \quad (6)$$

Finally, performing a range FFT on the above signal results in

$$S(t_a, f_r) = \text{rect}\left(\frac{t_a}{T_a}\right) \cdot \text{rect}\left(\frac{f_r}{\bar{B}_r}\right) \cdot \exp\left\{-j\frac{4\pi}{c}(\bar{f}_c + f_r)u\right\}. \quad (7)$$

Inserting $u = x_t \sin\theta + y_t \cos\theta$ into (7), we have

$$S(t_a, f_r) = \text{rect}\left(\frac{t_a}{T_a}\right) \cdot \text{rect}\left(\frac{f_r}{\bar{B}_r}\right) \cdot \exp\left\{-j\frac{4\pi}{c}(\bar{f}_c + f_r)(x_t \sin\theta + y_t \cos\theta)\right\}. \quad (8)$$

The signal after range preprocessing can be interpreted as discrete samples of the 2-D Fourier transform of the reflectivity function of the illuminated scene. However, the set of samples of Fourier domain lies on a polar raster imposed on an annulus. Therefore, in practical implementation of the image formation processing, to exploit the high efficiency of the 2-D IFFT to implement the 2-D inverse Fourier transform, a popular choice is to interpolate the 2-D polar-gridded samples into a rectangular grid for ultimate use in a 2-D IFFT image formation process.

*C. Polar Reformatting*

The polar format conversion is essentially a two-dimensional signal decoupling process, which is usually implemented by two one-dimensional resampling performed in range and azimuth respectively. The range resampling is a scaling transform of the range frequency as follows

$$f_r = \delta_r \bar{f}_r + \bar{f}_c(\delta_r - 1), \quad (9)$$

where $\delta_r = 1/\cos\theta$ is the scaling factor.

After range resampling, the signal becomes

$$S(t_a, \bar{f}_r) = \text{rect}\left(\frac{t_a}{T_a}\right) \cdot \text{rect}\left(\frac{\bar{f}_r}{\bar{B}_r}\right) \cdot \exp\left\{-j\frac{4\pi}{c}(\bar{f}_c + \bar{f}_r)(x_t \tan\theta + y_t)\right\}. \quad (10)$$

From the data collection geometry shown in Fig.1, it is clear that $\tan\theta = vt_a/R_e$. Therefore, (10) can also be expressed as

$$S(t_a, \bar{f}_r) = \text{rect}\left(\frac{t_a}{T_a}\right) \cdot \text{rect}\left(\frac{\bar{f}_r}{\bar{B}_r}\right) \cdot \exp\left\{-j\frac{4\pi}{c}(\bar{f}_c + \bar{f}_r)\left(x_t \frac{v}{R_e} t_a + y_t\right)\right\}. \quad (11)$$

The second step of polar reformatting is a range-frequency dependent azimuth resampling, which is mathematically described by a Keystone transform as follows

$$t_a = \frac{\bar{f}_c}{\bar{f}_c + \bar{f}_r} \bar{t}_a. \quad (12)$$

After azimuth resampling, the processed signal becomes

$$S(\bar{t}_a, \bar{f}_r) = \text{rect}\left(\frac{\bar{t}_a}{T_a}\right) \cdot \text{rect}\left(\frac{\bar{f}_r}{\bar{B}_r}\right) \cdot \exp\left\{-j\left[\frac{4\pi v}{\bar{\lambda} R_e} \bar{t}_a \cdot x_t + \frac{4\pi}{c}(\bar{f}_c + \bar{f}_r) \cdot y_t\right]\right\}, \quad (13)$$

where $\bar{\lambda} = c/\bar{f}_c$.

Now all the high-order terms and coupling terms in the phase history are eliminated, the signal becomes a 2-D sinusoid function.

*D. 2-D IFFT*

The final step is a 2-D IFFT. First, performing a range IFFT on (13) yields

$$S(\bar{t}_a, \bar{t}_r) = A \cdot \text{rect}\left(\frac{\bar{t}_a}{T_a}\right) \cdot \exp\left\{-j\left(\frac{4\pi v}{\bar{\lambda} R_e} \bar{t}_a \cdot x_t\right)\right\} \cdot \text{sinc}\left[\bar{B}_r\left(\bar{t}_r - \frac{2}{c} y_t\right)\right], \quad (14)$$

where $A = \exp\left\{-j\left(\frac{4\pi}{c} \bar{f}_c \cdot y_t\right)\right\}$ is a complex constant.

Then an azimuth IFFT produces the final focused image as

$$S(\bar{f}_a, \bar{t}_r) = A \cdot \text{sinc}\left[T_a\left(\bar{f}_a - \frac{2v}{\bar{\lambda} R_e} x_t\right)\right] \cdot \text{sinc}\left[\bar{B}_r\left(\bar{t}_r - \frac{2}{c} y_t\right)\right]. \quad (15)$$

Define $x = \frac{\bar{\lambda} R_e}{2v} \bar{f}_a$ and $y = \frac{c}{2} \bar{t}_r$, equation (15) can also be expressed as

$$S(x, y) = A \cdot \text{sinc}\left[\frac{2v T_a}{\bar{\lambda} R_e}(x - x_t)\right] \cdot \text{sinc}\left[\frac{2\bar{B}_r}{c}(y - y_t)\right]. \quad (16)$$

Now the point target is accurately focused at its true position in the orbital plane.

### III. EXTENDED SPHERICAL GEOMETRY ALGORITHM FOR STRIPMAP AND TOPS MODE SAR

In different SAR modes, the echo signals have the same phase history expression, but differ in the support region of the signal amplitude. If the signal discretization is not taken into account, the SGA can be directly applied to other modes without any modification. However, the actual collected echo signals are discretized in both dimensions. For discrete signals, Nyquist sampling is a prerequisite for joint processing between sampled values, e.g., resampling, FFT, etc. As the difference between the data in the different SAR modes is only in the azimuthal direction, we only need to discuss the azimuthal signal. In section A, we will analyze the Doppler history variation during SGA processing for different SAR modes under the continuous time assumption. In section B, azimuth sampling and aliasing problem in stripmap and TOPS modes will be discussed in detail. Finally, in section C, an alias-free image reconstruction algorithm is proposed for both stripmap and TOPS SAR data processing.

*A. Doppler History Variation during SGA Processing*

The beam steering geometry of three typical SAR modes is shown in the Fig.2. To simplify the analysis, the 2-D slant-range plane is adopted instead of the actual 3-D data collecting geometry. The radar platform moves at a speed $v$, and the radar beam is steered in different ways in different modes. These three different beam steering modes can be uniformly interpreted as the antenna beam always pointing to a fixed point (refer to as rotation center), and the difference between the different modes lies in the point it points to. In spotlight mode, the antenna is steered to point to a fixed point in the ground scene. In the stripmap mode, the pointing point is located at infinity, therefore the beam direction actually does not change during data collection. While for TOPS mode, there are usually two implementation methods, one is standard TOPS and the other is inverse TOPS. We only consider the standard TOPS mode in this paper. In standard TOPS mode, throughout the acquisition, the antenna is rotated from backward to forward, with the virtual rotation center at the other side of the flight trajectory with respect to the illuminated scene. Without loss of generality, assuming that the range from the scene center to the flight path is $R_{scene}$, the range from the virtual rotation center to the flight path in TOPS mode is $R_{centre}$.

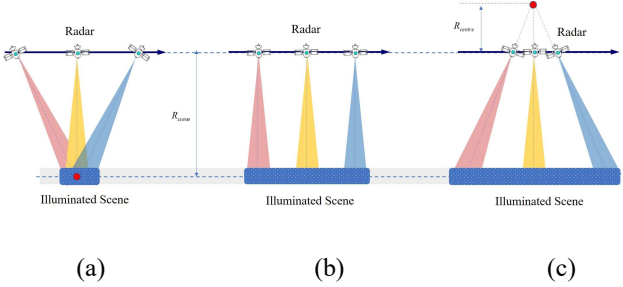

(a)  (b)  (c)

Fig.2. Beam steering in different SAR modes. (a) spotlight; (b) stripmap; (c) TOPS.

In different beam steering modes, the radar echo signal, after demodulation and range pulse compression, can be uniformly represented as

$$S(t_a, t_r) = A(t_a) \cdot \text{sinc}\left(B_r\left(t_r - \frac{2r}{c}\right)\right) \cdot \exp\left\{-j\frac{4\pi f_c}{c}r\right\}. \quad (17)$$

The difference in echo signals under different modes lies in the support zones of azimuth signals. Specifically,

$$A(t_a) = \text{rect}\left[(t_a / T_a)\right], \quad (18)$$

for spotlight mode,

$$A(t_a) = \text{rect}\left[(t_a - x_t / v) / T_u\right], \quad (19)$$

for stripmap, and

$$A(t_a) = \text{rect}\left[\left(t_a - \frac{R_{centre} x_t}{(R_{centre} + R_{scene})v}\right) / T_u\right], \quad (20)$$

for TOPS mode, where $T_a$ is the data acquisition time, and $T_u$ is the effective synthetic aperture time during which the radar illuminates the point scatterer.

From the SGA formulation, it is easy to obtain the Doppler history at different stages during SGA processing. Here, we are only concerned with the time-frequency relationships in three stages of the SGA processing process. The first is the time-frequency relationship in the phase history domain, the second is the time-frequency relationship before azimuth resampling, and the third is the time-frequency relationship before azimuth IFFT. The purpose of the phase history domain analysis here is not for the precise image formation processing, but only for the analysis of the sampling ambiguity in the following section. Therefore, only approximated time-frequency diagrams under the low-resolution SGA processing frame is provided, and the conclusion can be easily extended to the high resolution SGA case.

In the echo phase history domain, if the actual illumination time is not taken into account, the scatterers with different azimuthal positions in the scene will have the same Doppler history, and all of them can be approximated as linear FM signals, the difference is that the scattering points with different azimuthal positions have different zero Doppler times, whose time-frequency distributions are shown as the dotted lines in the figure. In practical situations, due to the limited width of the radar beam and the influence of the beam steering method, each scattering point is only illuminated by the radar for a limited time, so its phase history has only a limited support area. In the azimuth time domain, the entire echo signal support area is the data acquisition time, which is not fundamentally different for different modes. In the Doppler domain, the width of the signal support region, i.e., the instantaneous bandwidth, is approximately equal at each moment, and the difference between different modes lies in the varying instantaneous Doppler central frequency, as shown in Fig.3. In spotlight mode, the antenna is steered so that the beam is always pointing at the scene center. The instantaneous Doppler central frequency also changes linearly with time and at the same rate as the instantaneous Doppler frequency of the scattering points. The variation of the Doppler central frequency makes the total Doppler bandwidth of the signal in the spotlight mode much larger than the instantaneous Doppler bandwidth, as shown in Fig.3(a). In the stripmap mode, since the beam pointing is fixed, the Doppler central frequency is also fixed. For example, if the radar operates in sidelook stripmap mode, the Doppler central frequency is always zero, so the total signal bandwidth in this mode is equal to the signal instantaneous bandwidth, as shown in Fig.3(b). In the TOPS mode, the antenna look angle varies rapidly, and the Doppler central frequency also changes, and the change rate is greater than the change rate of the Doppler frequency of the scattering point. In this case, the total bandwidth of the azimuthal signal tends to be much larger than the instantaneous bandwidth of the signal, as shown in Fig.3(c).

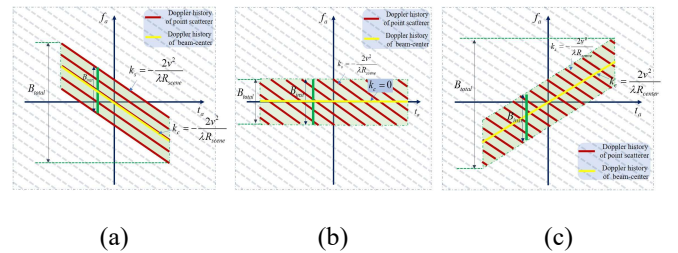

(a)  (b)  (c)

Fig.3. Time-frequency diagram of different SAR modes. (a) spotlight; (b) stripmap; (c) TOPS.

From (11), we can know that after processing by range preprocessing and range resampling, the azimuthal signal of each scatterer becomes a single-frequency signal, and the frequency value is linearly related with the azimuthal position of the scatterer. For example, assuming the azimuth position of the point target is $x_t$, its corresponding zero Doppler time is $x_t/v$. After these range processing, the original linear frequency modulation (LFM) signal becomes a single frequency signal, and the signal frequency is $f_t = -k_s x_t / v$. Therefore, by using IFFT, the target can be accurately focused in the frequency domain, and the position focused in the frequency domain is linearly related to the target azimuth position. If we do not consider the issue of sampling ambiguity, the above conclusions hold true regardless of whether it is spotlight, stripmap or TOPS mode. That is to say, if the signal sampling ambiguity problem is not considered, or if the pulse repetition frequency is high enough to satisfy the Nyquist sampling theorem in different modes, then the SGA can be directly applied to any mode without modifications. Fig. 4 shows the time-frequency diagram before and after SGA range processing in three different modes. The red line in the left figure represents the time-frequency relationship before range processing, and the blue line represents the time-frequency relationship after range processing. The right figure shows the focused image in the frequency domain.

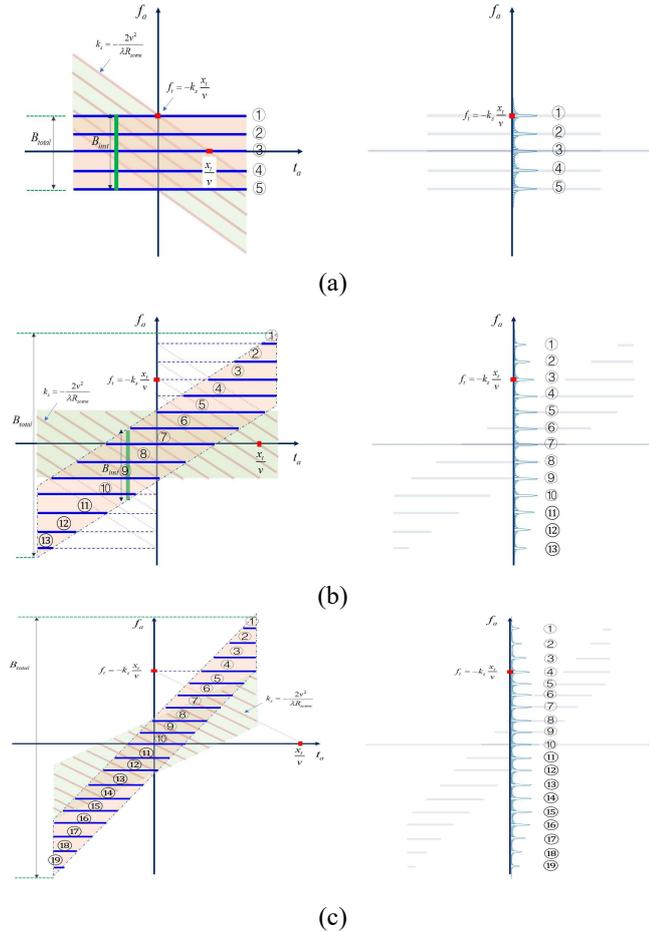

Fig.4. Time-frequency diagram after SGA processing and the focused image in the frequency domain. (a) spotlight; (b) stripmap; (c) TOPS.

Azimuth resampling in SGA is a two-dimensional decoupling of the signal to remove the residual linear range migration, which has no effect on the azimuth phase history. Therefore, the time-frequency relationship of the azimuth signal will not change after azimuth resampling.

B. *Azimuth Sampling and Aliasing*

In practical radar systems, azimuth signal acquisition is inherently discrete with the pulse repetition frequency (PRF) serving as the sampling rate. For spaceborne SAR system design, the PRF is typically configured marginally higher than the instantaneous azimuth signal bandwidth. Therefore, in the signal phase history domain, only the signal sampling in the stripmap mode meets the Nyquist sampling rate directly, while in the spotlight and TOPS modes, the PRF is often less than the total signal bandwidth. Nevertheless, in SGA formulation, the range processing, including the range preprocessing and the range resampling operation in the polar format conversion, is performed on each pulse echo independently, and these processing have no requirement for the PRF. Therefore, in the SGA processing, there is no need to care about the sampling ambiguity of the signal during range processing. However, both azimuth resampling and azimuth IFFT require joint processing of different azimuth pulse signals, so signal sampling in azimuth direction with Nyquist rate is an essential prerequisite for these processing operations.

According to the derivation in the previous section, we know that after range processing, the azimuth signal of each scatterer has changed into a single-frequency signal. At this time, although the instantaneous bandwidth at each moment has not changed, the total signal bandwidth has changed in different modes. In spotlight mode, the total bandwidth of the signal decreases and becomes equal to the instantaneous bandwidth, while in stripmap mode and TOPS mode, the total bandwidth of the signal increases, which is significantly greater than the instantaneous bandwidth. When the PRF is only slightly larger than the instantaneous bandwidth of the signal, sampling ambiguity in azimuth processing exists in both stripmap mode and TOPS mode. Therefore, in stripmap and TOPS mode, if the algorithm is not modified, both the azimuth resampling and azimuth IFFT in the SGA processing cannot yield correct results. For example, azimuth resampling cannot successfully decouple the 2-D signal to eliminate the residual linear range migration, resulting in 2-D defocused image. On the other hand, after azimuth IFFT, as shown in the Fig.5, the scatterers located in the scene edge will be aliasing in the image domain due to undersampling in the time domain. Therefore, in order to apply SGA algorithm to stripmap or TOPS mode SAR imaging processing, some necessary modifications are required.

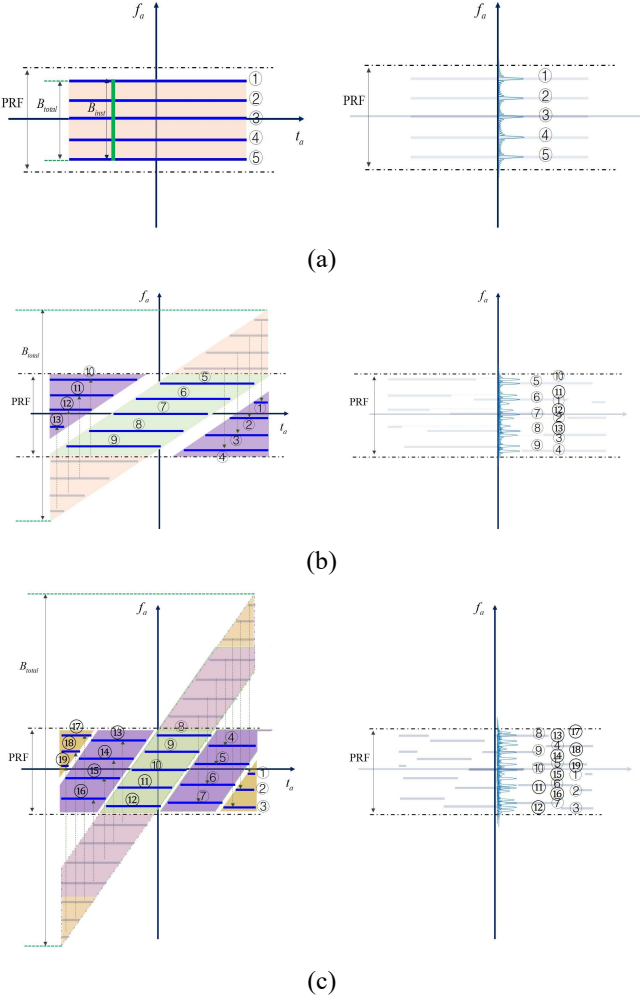

Fig.5. Time-frequency diagram and the produced SGA imagery in the frequency domain when considering limited PRF. (a) spotlight; (b) stripmap; (c) TOPS.

*C. Alias-free Image Reconstruction*

Let's first consider the sampling ambiguity problem during azimuthal resampling. Through the previous analysis, we know that the time-frequency distribution of the signal before azimuthal resampling is shown in Fig.5. Whether it is stripmap mode or TOPS mode, although the PRF is always larger than the instantaneous bandwidth at any moment, the total bandwidth of the signal is often much larger than the PRF because the instantaneous Doppler central frequency varies with the time. In order to avoid the sampling ambiguity, an additional phase correction on the signal is required before the azimuthal resampling, which eliminates the variation of the instantaneous Doppler central frequency to ensure that the signal sampling is not ambiguous.

To simplify the analysis, assume that the instantaneous Doppler central frequency varies linearly with time at a rate of $k_t$ ($k_t = 2v^2/(\lambda R_{scene})$ for stripmap mode, and $k_t = 2v^2/(\lambda R_{scene}) + 2v^2/(\lambda R_{centre})$ for TOPS mode). Therefore, in order to compensate for the instantaneous Doppler central frequency variation, the following azimuth compensation function should be multiplied on the data

$$H_1(t_a) = \exp\{-j\pi k_t t_a^2\} \ . \quad (21)$$

The corrected signal then can be represented as

$$S(t_a, \bar{f}_r) = A(t_a) \cdot \text{rect}\left(\frac{\bar{f}_r}{\bar{B}_r}\right) \cdot \exp\left\{-j\left[\frac{4\pi}{c}(\bar{f}_c + \bar{f}_r)\left(x_t \frac{v}{R_e} t_a + y_t\right) + \pi k_t t_a^2\right]\right\}. \quad (22)$$

After compensation, the instantaneous Doppler central frequency always remains zero, and the total bandwidth of the azimuth signal is equal to the instantaneous bandwidth of the signal, as shown in the Fig.6. At this time, the signal sampling satisfies the Nyquist sampling theorem, so the subsequent azimuth interpolation and Fourier transform processing can be correctly performed.

First, let's look at the azimuth resampling process. Azimuth interpolation processing is essentially a Keystone transform, i.e. $t_a = \frac{\bar{f}_c}{\bar{f}_c + \bar{f}_r} \bar{t}_a$. Therefore, the signal after azimuth resampling can be expressed as

$$S(\bar{t}_a, \bar{f}_r) = A(\bar{t}_a) \cdot \text{rect}\left(\frac{\bar{f}_r}{\bar{B}_r}\right) \cdot \exp\left\{-j\left[\frac{4\pi}{c}(\bar{f}_c + \bar{f}_r) \cdot y_t \right.\right.$$
$$\left.\left. + \frac{4\pi \bar{f}_c v}{c R_e} \bar{t}_a \cdot x_t + \pi k_t \left(\frac{\bar{f}_c}{\bar{f}_c + \bar{f}_r}\right)^2 \bar{t}_a^2\right]\right\} . \quad (23)$$

From the above equation, it can be seen that after the azimuthal resampling, the original range-azimuth coupling of the 2-D signal has been decoupled, but a new coupling term has appeared. In order to avoid the emergence of new coupling terms, the compensation function in Eq. (21) should be modified as

$$H_1(t_a) = \exp\left\{-j\pi k_t \left(\frac{\bar{f}_c + \bar{f}_r}{\bar{f}_c}\right)^2 t_a^2\right\} \ . \quad (24)$$

In this case, the signal after azimuth resampling becomes

$$S(\bar{t}_a, \bar{f}_r) = A(\bar{t}_a) \cdot \text{rect}\left(\frac{\bar{f}_r}{\bar{B}_r}\right) \cdot \exp\left\{-j\left[\frac{4\pi}{c}(\bar{f}_c + \bar{f}_r) \cdot y_t + \frac{4\pi \bar{f}_c v}{c R_e} \bar{t}_a \cdot x_t + \pi k_t \bar{t}_a^2\right]\right\}. \quad (25)$$

At this point, the two-dimensional coupling of the signal is completely removed, but there is still an additional quadratic phase function in the azimuth direction, which means that the azimuth signal becomes an LFM signal again. In order to use the original SGA algorithm for azimuth imaging, it is necessary to remove this quadratic phase term. However, removing this quadratic phase function will inevitably result in a change in the instantaneous Doppler central frequency, thus increasing the signal bandwidth, with the consequence that the final azimuthal IFFT imaging will produce aliased image for the edge scatterers.

In order to avoid image aliasing, a feasible method is to perform appropriate oversampling during azimuth resampling to increase the azimuthal sampling rate. As long as the increased sampling rate is large enough, a direct azimuthal IFFT can still obtain an unaliased image although the total bandwidth is increased after removing the quadratic term in Eq. (25). The cost of this solution is a dramatic increase in the amount of data to be processed. An alternative approach that does not require a significant increase in the amount of data is to use matched filtering instead of spectrum analysis to obtain the focused image. After the azimuth resampling, the azimuthal signal becomes an LFM signal again if the quadratic phase term introduced earlier is not compensated for. For the convenience of analysis, we combine the quadratic phase term and the linear phase term in the azimuth signal. For stripmap mode, the signal becomes

$$S(\bar{t}_a, \bar{f}_r) = \text{rect}\left[\left(\bar{t}_a - \frac{x_t}{v}\right)/T_u\right] \cdot \text{rect}\left(\frac{\bar{f}_r}{\bar{B}_r}\right) \cdot \exp\left\{-j\left[\frac{4\pi}{c}(\bar{f}_c + \bar{f}_r) \cdot y_t + \pi k_t \left(\bar{t}_a - \frac{x_t}{v}\right)^2\right]\right\}, \quad (26)$$

where $k_t = 2v^2/(\lambda R_{scene})$ and the nonessential amplitude factor is eliminated for clarity.

For TOPS mode, the signal can be rewritten as

$$S(\bar{t}_a, \bar{f}_r) = \text{rect}\left[\left(\bar{t}_a - \frac{R_{centre} x_t}{(R_{scene} + R_{centre})v}\right)/T_u\right] \cdot \text{rect}\left(\frac{\bar{f}_r}{\bar{B}_r}\right) \cdot \exp\left\{-j\left[\frac{4\pi}{c}(\bar{f}_c + \bar{f}_r) \cdot y_t + \pi k_t \left(\bar{t}_a - \frac{R_{centre} x_t}{(R_{scene} + R_{centre})v}\right)^2\right]\right\}, \quad (27)$$

where $k_t = 2v^2/(\lambda R_{scene}) + 2v^2/(\lambda R_{centre})$.

At this point, whether in stripmap or TOPS mode, the azimuth signal is a standard LFM signal, and the instantaneous Doppler central frequency is always zero, as shown in Fig.6. In this case, if matched filtering is used to realize azimuthal compression, there is no need to oversample the data during azimuth resampling.

The matched filter processing consists of three steps. The first operation is an azimuthal FFT.

For stripmap mode, the signal after azimuth FFT becomes

$$S(\bar{f}_a, \bar{f}_r) = \text{rect}\left(\frac{\bar{f}_a}{k_t T_u}\right) \cdot \text{rect}\left(\frac{\bar{f}_r}{\bar{B}_r}\right) \cdot \exp\left\{-j\left[\frac{4\pi}{c}(\bar{f}_c + \bar{f}_r) \cdot y_t\right]\right\} \cdot \exp\left\{j\left(\pi \frac{\bar{f}_a^2}{k_t} + 2\pi \bar{f}_a \frac{x_t}{v}\right)\right\}. \quad (28)$$

While for TOPS mode, the signal after azimuth FFT becomes

$$S(\bar{f}_a, \bar{f}_r) = \text{rect}\left(\frac{\bar{f}_a}{k_t T_u}\right) \cdot \text{rect}\left(\frac{\bar{f}_r}{\bar{B}_r}\right) \cdot \exp\left\{-j\left[\frac{4\pi}{c}(\bar{f}_c + \bar{f}_r) \cdot y_t\right]\right\} \cdot \exp\left\{j\left[\pi \frac{\bar{f}_a^2}{k_t} + 2\pi \bar{f}_a \frac{R_{centre} x_t}{(R_{scene} + R_{centre})v}\right]\right\}. \quad (29)$$

The second step is to multiply the signal by the following reference function

$$H(\bar{f}_a) = \exp\left\{-j\left(\pi \frac{\bar{f}_a^2}{k_t}\right)\right\}. \quad (30)$$

For stripmap mode, we obtain

$$S(\bar{f}_a, \bar{f}_r) = \text{rect}\left(\frac{\bar{f}_a}{k_t T_u}\right) \cdot \text{rect}\left(\frac{\bar{f}_r}{\bar{B}_r}\right) \cdot \exp\left\{-j\left[\frac{4\pi}{c}(\bar{f}_c + \bar{f}_r) \cdot y_t\right]\right\} \cdot \exp\left\{j\left(2\pi \bar{f}_a \frac{x_t}{v}\right)\right\}. \quad (31)$$

For TOPS mode, the signal becomes

$$S(\bar{f}_a, \bar{f}_r) = \text{rect}\left(\frac{\bar{f}_a}{k_t T_u}\right) \cdot \text{rect}\left(\frac{\bar{f}_r}{\bar{B}_r}\right) \cdot \exp\left\{-j\left[\frac{4\pi}{c}(\bar{f}_c + \bar{f}_r) \cdot y_t\right]\right\} \cdot \exp\left\{j\left[2\pi \bar{f}_a \frac{R_{centre} x_t}{(R_{scene} + R_{centre})v}\right]\right\}. \quad (32)$$

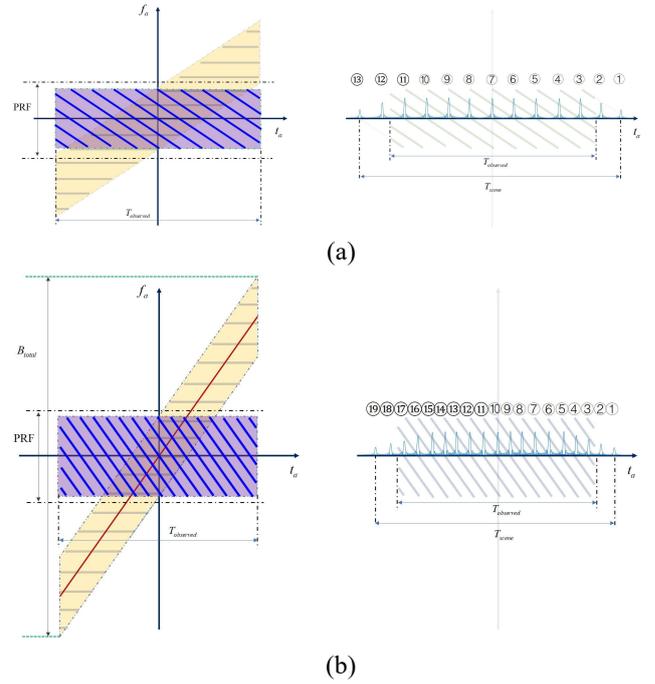

(a)

(b)

Fig.6. Time-frequency diagram after instantaneous doppler central frequency elimination and the produced SGA imagery in the azimuth time domain by match filtering. (a) stripmap; (b) TOPS.

The third step is an azimuth IFFT, which can produce the azimuth focused image

For stripmap mode, the azimuth focused imagery is

$$S(\bar{t}_a, \bar{f}_r) = \text{rect}\left(\frac{\bar{f}_r}{\bar{B}_r}\right) \cdot \exp\left\{-j\left[\frac{4\pi}{c}(\bar{f}_c + \bar{f}_r) \cdot y_t\right]\right\} \cdot \text{sinc}\left[k_t T_u \left(\bar{t}_a - \frac{x_t}{v}\right)\right]. \quad (33)$$

For TOPS mode, the azimuth focused imagery is

$$S(\bar{t}_a, \bar{f}_r) = \text{rect}\left(\frac{\bar{f}_r}{\bar{B}_r}\right) \cdot \exp\left\{-j\left[\frac{4\pi}{c}(\bar{f}_c + \bar{f}_r) \cdot y_t\right]\right\} \cdot \text{sinc}\left[k_t T_u \left(\bar{t}_a - \frac{R_{centre} x_t}{(R_{scene} + R_{centre})v}\right)\right]. \quad (34)$$

Finally, a range IFFT can produce a full focused imagery. For stripmap mode, the full focused imagery is

$$S(\bar{t}_a, \bar{t}_r) = \text{sinc}\left[k_t T_u\left(\bar{t}_a - \frac{x_t}{v}\right)\right] \cdot \text{sinc}\left[\bar{B}_r\left(\bar{t}_r - \frac{2y_t}{c}\right)\right]. \quad (35)$$

While for TOPS mode, the full focused imagery is

$$S(\bar{t}_a, \bar{t}_r) = \text{sinc}\left[k_t T_u\left(\bar{t}_a - \frac{R_{centre} x_t}{(R_{scene} + R_{centre})v}\right)\right] \cdot \text{sinc}\left[\bar{B}_r\left(\bar{t}_r - \frac{2y_t}{c}\right)\right]. \quad (36)$$

Define $x = v\bar{t}_a, y = c\bar{t}_r/2$ to convert the 2-D time-domain coordinates to 2-D space-domain coordinates, Eq.(35) and Eq.(36) can also be expressed as

$$S(x, y) = \text{sinc}\left[\frac{k_t T_u}{v}(x - x_t)\right] \cdot \text{sinc}\left[\frac{2\bar{B}_r}{c}(y - y_t)\right], \quad (37)$$

$$S(x, y) = \text{sinc}\left[\frac{k_t T_u}{v}\left(x - \frac{R_{centre}}{(R_{scene} + R_{centre})}x_t\right)\right] \cdot \text{sinc}\left[\frac{2\bar{B}_r}{c}(y - y_t)\right]. \quad (38)$$

In summary, the flow chart of the proposed extended SGA for processing stripmap and TOPS SAR data is shown in Fig.7.

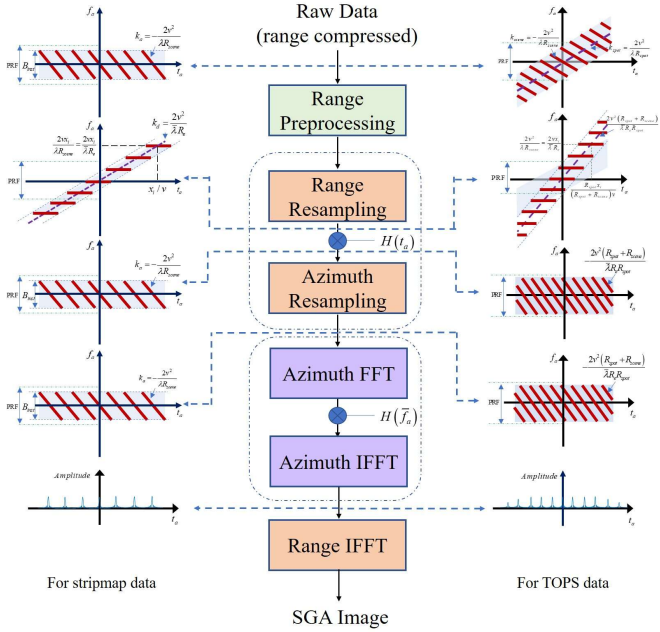

Fig.7. Flowchart of the proposed extended SGA.

## IV. EXPERIMENTAL RESULTS

In order to validate the proposed algorithm, processing results of real SAR data are presented. The measured data was collected by a Chinese commercial spaceborne SAR satellite Chaohu-1. The radar platform of Chaohu-1 is a low earth orbit (LEO) satellite with an orbit height of 500km. The payload is a C-band synthetic aperture radar which has different beam steering modes and different resolution. In this example, both stripmap and TOPS mode data are processed by using the algorithm presented in this article. Some main orbit and radar parameters are shown in Table I. In order to demonstrate the necessity of improving the SGA algorithm in stripmap and TOPS modes, the classical SGA algorithm is also used to process the data in both modes.

TABLE I. SYSTEM PARAMETERS

| Parameter | Stripmap | TOPS |
|---|---|---|
| Carrier Frequency | 5.4GHz | 5.4GHz |
| Range Bandwidth | 200MHz | 40MHz |
| Pulse Duration | 24us | 22us |
| Pulse Repeat Frequency | 3900Hz | 4965Hz |
| Coherent Time | 4s | 0.8s |
| Orbit Altitude | 532km | 544km |
| Reference Range | 597km | 646km |
| Nominal Resolution | 1.5m*1.5m | 7m*10m |
| Scene Size(range*azimuth) | 20km*30km | 36km*38km |

### A. Results of Stripmap SAR Data

The dimension of the stripmap mode data is 16384 * 16384. The theoretical range resolution is 1.5m, and the theoretical azimuth resolution is 1.5m. The size of the illuminated scene during data acquisition is approximately 20km*30km. Firstly, the data is processed using the classical SGA algorithm, and the imaging results are shown in the Fig.8(a) and Fig.9(a), where Fig.8(a) is the range compressed image, and Fig.9(a) is the full compressed image. From the Fig.8(a), it can be seen that the range migration of some scatterers has been completely corrected, but there are also some scatterers that still have residual linear range migration. The reason is that the range migration of scatterers at the azimuth edge of the scene cannot be fully corrected during azimuth resampling due to the undersampling of the data. On the other hand, as can be seen from Fig.9(a), the scatterers at the azimuth edge of the scene are aliased in the produced image due to undersampling during the final IFFT processing. Also, these scatterers appear 2-D defocus due to the incomplete range migration correction.

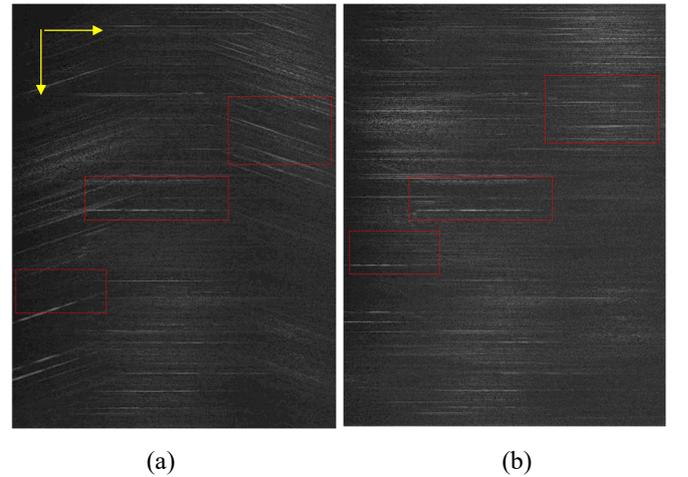

(a)          (b)

Fig.8. The range compressed image after classic SGA and the extended SGA processing.

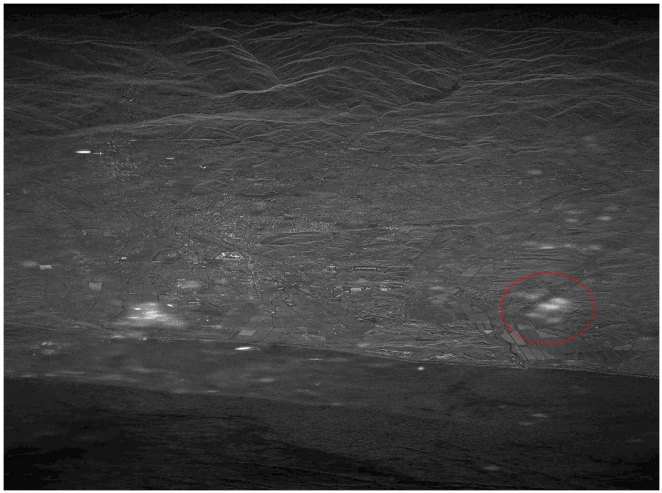

(a)

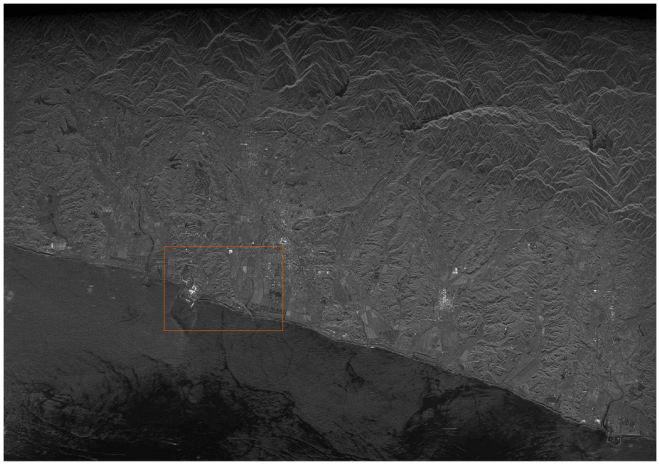

(b)

Fig.9. The full compressed image after classic SGA and the extended SGA processing in stripmap mode.

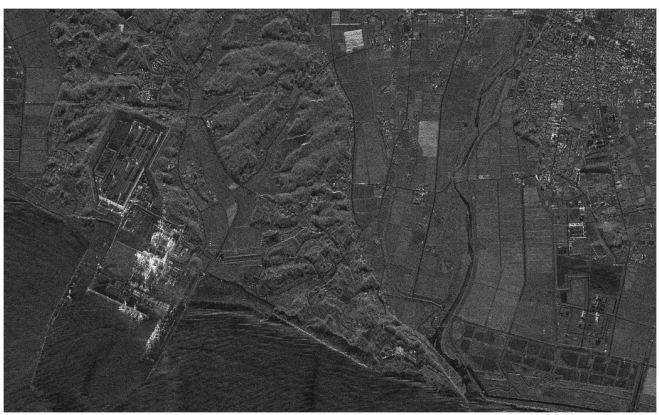

Fig.10. An enlarged local area in Fig.9.

Fig.8(b) and Fig.9(b) show the processing results by the proposed approach. Fig.8(b) is the range-compressed image. It is clear that range migration effect is completely eliminated for all scatterers. Fig.9(b) is the full-compressed image. To see more clearly the focus quality, an enlarged local subimage in Fig. 9(b) is also provided in Fig.10. From these figures, we can see that all the scatterers, not just those in the scene centre but also those at the edges of the scene, are all well focused. Both the residual range migration effect and image aliasing effect faced by the traditional SGA have been effectively eliminated by the proposed algorithm.

*B. Results of TOPS SAR Data*

The second raw data set is collected by TOPS mode. During the data collection, the range of rotation center from the flight path is 150km, the data acquiring time is 0.8s, which corresponding to a synthetic aperture length of 6135m. During the beam steering, the illuminated scene dimension in azimuth is 38km. The theoretical ground range resolution is 7m, and the theoretical azimuth resolution is 8m. Similarly, the classic SGA is firstly applied to process the data and the processing results are shown in Fig.11(a). Compared with the stripmap SAR imagery shown in Fig.9(a), the TOPS SAR imagery produced by classic SGA algorithm suffer from more severe aliasing effects due to the larger imaging scene. Nevertheless, after processing by the proposed algorithm, as shown in Fig.11(b), all scatterers in the illuminated scene are accurately focused at their true positions in the orbital plane.

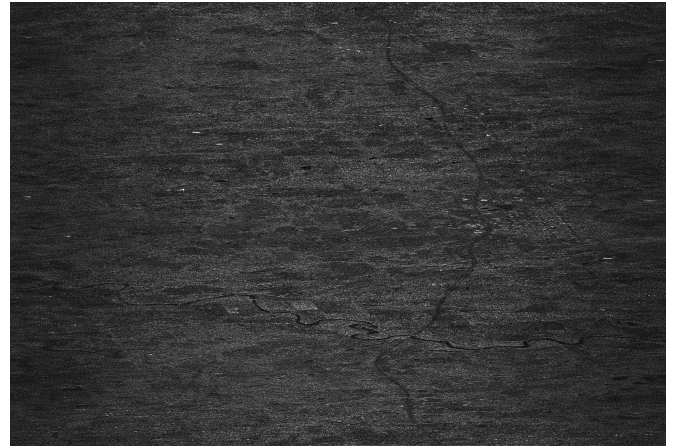

(a)

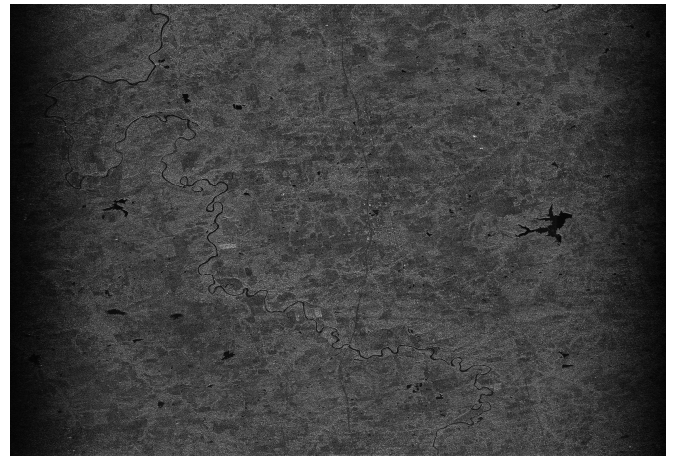

(b)

Fig.11. The full compressed image after classic SGA and the extended SGA processing in TOPS mode.

## V. Conclusion

In this paper, we extended the traditional SGA to process stripmap and TOPS SAR data. Compared to the original algorithm, the main difference of the improved algorithm lies in azimuth resampling and azimuth compression. Firstly, in order to avoid undersampling during azimuth resampling in stripmap and TOPS modes, an instantaneous Doppler centroid removal process was added before azimuth interpolation processing. Secondly, the spectral analysis method used for the final step of azimuth compression in the original SGA has been replaced with a new matched filtering processing, which can avoid image aliasing in azimuth direction. The extended algorithm greatly expands the application scenarios of the SGA algorithm.

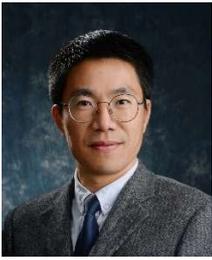 **Xinhua Mao** (Member, IEEE) was born in Hunan, China. He received the B.S. and Ph.D. degrees from the Nanjing University of Aeronautics and Astronautics (NUAA), Nanjing, China, in 2003 and 2009, respectively, all in electronic engineering.

In 2009, he joined the Department of Electronic Engineering, NUAA, where he is currently a Full Professor with the Key Laboratory of Radar Imaging and Microwave Photonics. He was a Visiting Scholar with Villanova University, Villanova, PA, USA, in 2013, and the University of Leicester, Leicester, U.K., from 2018 to 2019. His research interests include radar imaging, and ground moving target indication (GMTI), inverse problems. He has developed algorithms for several operational airborne SAR systems.

Dr. Mao was a recipient of the one National Science and Technology Progress Award in 2019 and three National Defense Technology Awards in 2007, 2015, and 2018, in China. He received the Best Paper Award in the 5th Asia-Pacific Synthetic Aperture Radar Conference in 2015.